\documentclass[aps,floatfix,prd,twocolumn,a4paper,10pt,citeautoscript,
preprintnumbers,superscriptaddress,showpacs,longbibliography]{revtex4-1} 

\usepackage[english]{babel} 	
\usepackage{amsmath}        	
\usepackage{amssymb}        	
\usepackage{bm}					
\usepackage{bbm}            	
\usepackage{empheq}
\usepackage{array}		

\usepackage{graphicx}       
\usepackage{graphics}       
\DeclareGraphicsExtensions{.jpg,.pdf} 
\usepackage{hyperref}
\hypersetup{colorlinks=true, pdfstartview=FitV, 
linkcolor=blue, citecolor=blue, urlcolor=blue}

\renewcommand{\theequation}{\arabic{equation}}
\newcommand{\Equation}[2]{\begin{equation}\label{#1}#2\end{equation}}
\newcommand{\Align}[2]{\begin{align}\label{#1}#2\end{align}}
\newcommand{\SubAlign}[2]{\begin{subequations}\label{#1}\begin{align}#2\end{align}\end{subequations}}

\newcommand{\bs}{\boldsymbol}
\newcommand{\Figref}[1]{Fig.~\ref{#1}}
\newcommand{\Eqref}[1]{\eqref{#1}}
\newcommand{\Exp}[1]{\text{e}^{#1}}
\renewcommand\Re{\mathrm{Re}}

\newcommand{\Grad}{{\bs\nabla}}

\newcommand{\Curl}{{\bs\nabla}\times}

\newcommand{\A}{{\bs A}}
\newcommand{\B}{{\bs B}}
\newcommand{\D}{{\bs \Pi}}
\newcommand{\F}{\mathcal{F}}

\begin{document}
\title{Field-induced coexistence of \texorpdfstring{$s_{++}$}{s++} 
and \texorpdfstring{$s_{\pm}$}{s+-} superconducting states in dirty multiband 
superconductors}
\author{Julien~Garaud} 
\affiliation{Department of Physics, KTH-Royal Institute of Technology, 
Stockholm, SE-10691 Sweden} 
\affiliation{
Laboratoire de Math\'{e}matiques et Physique Th\'{e}orique CNRS/UMR 7350, \\ 
F\'{e}d\'{e}ration Denis Poisson FR2964, 
Universit\'{e} de Tours, Parc de Grandmont, 37200 Tours, France}
\author{Alberto~Corticelli}  
\affiliation{Department of Physics, KTH-Royal Institute of Technology, 
Stockholm, SE-10691 Sweden} 
\author{Mihail~Silaev}	
\affiliation{Department of Physics and Nanoscience Center, 
University of Jyv\"askyl\"a, P.O. Box 35 (YFL), FI-40014 
University of Jyv\"askyl\"a, Finland}
\author{Egor~Babaev} 
\affiliation{Department of Physics, KTH-Royal Institute of Technology, 
Stockholm, SE-10691 Sweden} 

\begin{abstract} 

In multiband systems, such as iron-based superconductors, the superconducting 
states with locking and anti-locking of the interband phase differences, 
are usually considered as mutually exclusive. For example, a dirty two-band 
system with interband impurity scattering undergoes a sharp crossover between 
the $s_{\pm}$ state (which favors phase anti locking) and the $s_{++}$ state 
(which favors phase locking). 
We discuss here that the situation can be much more complex in the presence 
of an external field or superconducting currents. In an external applied 
magnetic field, dirty two-band superconductors do not feature a sharp 
$s_{\pm}\to s_{++}$ crossover but rather a washed-out crossover to a finite 
region in the parameter space where both $s_{\pm}$ and $s_{++}$ states can 
coexist for example as a lattice or a microemulsion of inclusions of different 
states.  
The current-carrying regions such as the regions near vortex cores can exhibit 
an $s_\pm$ state while it is the $s_{++}$ state that is favored in the bulk.
This coexistence of both states can even be realized in the Meissner state 
at the domain's boundaries featuring Meissner currents. We demonstrate that 
there is a magnetic-field-driven crossover between the pure $s_{\pm}$ and 
the $s_{++}$ states.

\end{abstract}

\date{\today}
\maketitle

\section{Introduction}

It is widely accepted that in iron-based superconductors, the pairing between 
electrons is produced by interband electron-electron repulsion 
\cite{Mazin.Singh.ea:08,Chubukov.Efremov.ea:08,Hirschfeld.Korshunov.ea:11}. 
In such a situation, the superconducting state, which features a sign inversion 
between the two $s$-wave gap functions is called $s_\pm$, in contrast to the 
$s_{++}$ state, which has no sign inversion. The presence of disorder is known 
to potentially lead to a crossover from the $s_{\pm}$ to the $s_{++}$ state 
\cite{Efremov.Korshunov.ea:11}. In the absence of an external magnetic field, 
the crossover is sharp and has little thermodynamic features.
It was however recently demonstrated that the crossover line is accompanied 
by a non trivial transition in the core structure of vortices 
\cite{Garaud.Silaev.ea:17a}. 
More precisely the vortices, in the vicinity of the crossover line, can acquire 
a circular nodal line around the singular point in one of the superconducting 
components \cite{Garaud.Silaev.ea:17a}. This singular nodal line, which in 
three dimensions extends to a cylindrical nodal surface surrounding the vortex 
line, results in the formation of a peculiar ``moat"-like profile in the 
subdominant component of the superconducting gap. As a result, the inner 
region of the vortex core shows a $\pi$ relative phase between the gaps while 
it is zero in the outer region. This means that these moat-core vortices consist 
of an $s_\pm$ phase inclusion in the vortex core, which is separated from the 
bulk $s_{++}$ phase by the nodal line. 
Here we investigate the consequences of that physics on the phase diagram of 
such dirty two-band superconductors, in an external magnetic field. In a low 
applied external field the lattices and liquids of such moat-core vortices 
represent a phase coexistence or a mircoemulsion of such $s_\pm$ inclusions 
inside the bulk $s_{++}$ state. At elevated fields this results in a field-induced 
transition between the $s_\pm$ and $s_{++}$ states.

We start our investigation, in Sec.~\ref{Sec:GL}, by deriving a two-band 
Ginzburg-Landau model (including the gradient terms) from the microscopic 
Usadel theory of dirty two-band superconductors and discuss the essential 
properties of the phase diagram. 
Next, Sec.~\ref{Sec:Vortices} is devoted to the investigation of the physical 
properties of the elementary topological excitations (the vortices) and their 
core structure in that model. There, we show that across the $s_\pm/s_{++}$ 
crossover line there is a structural transition in the core of topological 
excitations, and we further construct the diagram where such solutions 
occur, as a function of the system's microscopic parameters. 
In Sec.~\ref{Sec:Magnetization} we investigate the consequences for the 
phase diagram in an applied external magnetic field, and finally our 
conclusions are presented in Sec.~\ref{Sec:Conclusion}.

\section{Theoretical framework}\label{Sec:GL}

Within a weak-coupling approximation, two-band superconductors with a high 
concentration of impurities can be described by a system of two Usadel 
equations, coupled by interband impurity-scattering terms (see e.g.
\cite{Gurevich:03}):
\Align{Eq:Usadel}{
\omega_n f_i &= \frac{D_i}{2} \big(g_i\D^2 f_i - f_i \nabla^2 g_i\big) 	
			 +  \Delta_i g_i									\nonumber \\
 			 &~~~+\sum_{j\neq i}\gamma_{ij} ( g_if_j - g_jf_i) 	\,.
}
Here $\omega_n = (2n+1)\pi T $ (with $n\in \mathbb{Z}$) are the Matsubara 
frequencies. $T$ is the temperature, $D_i$ are the electron diffusivities, 
and $\gamma_{ij}$ are the interband scattering rates. 
The quasiclassical propagators $f_i$ and $g_i$ are respectively the anomalous 
and normal Green's functions in each band, which obey the normalization 
condition $|f_i|^2 + g_i^2 =1$. The components of the order parameter 
$\Delta_j=|\Delta_j|e^{i\theta_j}$ are determined by the self-consistency
equations
\Equation{Eq:SelfConsistency}{
 \Delta_i =2\pi T  \sum_{n=0}^{N_d} 
 \sum_{j} \lambda_{ij} f_{j} (\omega_n),
}
for the Green's functions that satisfy Eq.~\Eqref{Eq:Usadel}. Here,  
$N_d=\Omega_d/(2\pi T)$ is the summation cut-off at Debye frequency $\Omega_d$. 
The diagonal elements $\lambda_{ii}$ of the coupling matrix $\hat{\lambda}$ 
in the self-consistency equation \Eqref{Eq:SelfConsistency} describe the 
intraband pairing, while the interband interaction is determined by the 
off-diagonal terms $\lambda_{ij}$ ($j\neq i$). The interband coupling parameters 
and impurity scattering amplitudes satisfy the symmetry relation \cite{Gurevich:03}
\Equation{Eq:Symmetry}{
 \lambda_{ij}= - \lambda_J/N_i ~~\text{with}~~j\neq i,
 ~~\text{and}~~\gamma_{ij}=\Gamma N_j \,,
} 
where $\lambda_J,\Gamma>0$ and $N_{1,2}$ are the partial densities of states 
in the two-bands.  

In order to investigate the physical properties of dirty two-band superconductors 
in an external field, we consider a Ginzburg-Landau (GL) model that is derived from 
the microscopic Usadel theory of dirty superconductors. The Ginzburg-Landau 
free energy functional for multiband models is obtained as an expansion in 
several small parameters: small gaps and gradients (not to be confused with a 
symmetry-based GL expansion that uses a single small parameter $\tau$; see also 
remark \footnote{For the question of formal validity of the multiband expansion 
see the discussion of the clean case \cite{Silaev.Babaev:12}}). \nocite{Silaev.Babaev:12}
The resulting expression, including gradient terms, reads as 
\cite{Garaud.Silaev.ea:17a}:
\SubAlign{Eq:FreeEnergy}{
\frac{\F}{\F_0} =&
\sum_{j=1}^2\Big\{
 	\frac{k_{jj}}{2}\left|\D\Delta_j \right|^2
 	+a_{jj}|\Delta_j|^2+\frac{b_{jj}}{2}|\Delta_j|^4\Big\}  
 	\label{Eq:FreeEnergy:Self}	\\
   &+\frac{k_{12}}{2}
 	\Big((\D\Delta_1)^*\D\Delta_2+(\D\Delta_2)^*\D\Delta_1 \Big)
	\label{Eq:FreeEnergy:Mixed}	\\
   &+2\left(a_{12}+c_{11}|\Delta_1|^2+c_{22}|\Delta_2|^2\right)
 	 \Re\big(\Delta_1^*\Delta_2\big)
	\label{Eq:FreeEnergy:Interaction1}	\\
   &+\left(b_{12}+c_{12}\cos2\theta_{12}\right)|\Delta_1|^2|\Delta_2|^2    
   +\frac{\B^2}{2}
	\label{Eq:FreeEnergy:Interaction2}	\,.
}
Here, the complex fields $\Delta_j=|\Delta_j| e^{i\theta_j}$ represent the 
superconducting gaps in the different bands, and $\theta_{12}=\theta_2-\theta_1$
stands for the relative phase between them. The two gaps in the different 
bands are electromagnetically coupled by the vector potential $\A$ of the 
magnetic field $\B=\bs\nabla\times\A$, through the gauge derivative 
$\D\equiv\Grad+iq\A$, and the coupling constant $q$ parametrizes the 
penetration depth of the magnetic field.

The coefficients of the Ginzburg-Landau functional $a_{ij}$, $b_{ij}$, $c_{ij}$ 
and $k_{ij}$ are calculated from a given set of input parameters $\lambda_{ij}$, 
$D_i$, $T$ and $\Gamma$ of the microscopic self-consistent equation. First, 
the coefficients of gradient terms are given by \cite{Garaud.Silaev.ea:17a}
\SubAlign{Eq:GLparameters:K}{
 k_{ii} &= 2\pi T N_i \sum_{n=0}^{N_d} 
 \frac{ D_i (\omega_n + \gamma_{ji})^2 + \gamma_{ij}\gamma_{ji} D_j }
 {\omega^2_n(\omega_n+\gamma_{ij}+\gamma_{ji})^2} \\
 k_{ij} &= 2\pi T N_i\gamma_{ij} \sum_{n=0}^{N_d} 
 \frac{ D_i (\omega_n + \gamma_{ji}) + D_j (\omega_n + \gamma_{ij}) }
 {\omega^2_n(\omega_n+\gamma_{ij}+\gamma_{ji})^2}	\,,
}
with $j\neq i$. Next, the coefficients of the potential terms are
\SubAlign{Eq:GLparameters:A}{
a_{ii}&= \frac{N_i\lambda_{jj}}{\mathrm{det}(\hat{\lambda})}
-2\pi T \sum_{n=0}^{N_d}
\frac{(\omega_n+\gamma_{ji}) N_i}{\omega_n(\omega_n+\gamma_{ij}+\gamma_{ji})}
\,, \\
a_{ij}&=-\frac{N_i\lambda_{ij}}{\mathrm{det}(\hat{\lambda})}
-2\pi T \sum_{n=0}^{N_d}
\frac{ \gamma_{ij} N_i}{\omega_n(\omega_n+\gamma_{ij}+\gamma_{ji})}
\,. 
}
The other parameters read as 
\SubAlign{Eq:GLparameters:B}{
b_{ii}&= \pi T N_i\sum_{n=0}^{N_d}
\frac{(\omega_n+\gamma_{ji})^4 }{\omega_n^3(\omega_n+\gamma_{ij}+\gamma_{ji})^4}
\\ &~+
\pi T N_i\sum_{n=0}^{N_d}
\frac{\gamma_{ij}(\omega_n+\gamma_{ji})
(\omega_n^2+3\omega_n\gamma_{ji}+\gamma_{ji}^2) }
{\omega_n^3(\omega_n+\gamma_{ij}+\gamma_{ji})^4}
\,,\nonumber \\
b_{ij}&= -\pi T N_i\sum_{n=0}^{N_d}
\frac{\gamma_{ij} }{(\omega_n+\gamma_{ij}+\gamma_{ji})^4}
\\ +
\pi &T N_i\sum_{n=0}^{N_d}
\frac{\gamma_{ij}(\gamma_{ij}+\gamma_{ji})(\omega_n(\gamma_{ij}+\gamma_{ji})
+2\gamma_{ij}\gamma_{ji})}
{\omega_n^3(\omega_n+\gamma_{ij}+\gamma_{ji})^4}
\,,\nonumber
}
and 
\SubAlign{Eq:GLparameters:C}{
c_{ii}&=\pi T N_i \\ 
&\sum_{n=0}^{N_d}
\frac{  \gamma_{ij}(\omega_n+\gamma_{ji})
(\omega_n^2 +(\omega_n+\gamma_{ji})(\gamma_{ij}+\gamma_{ji})) }
{\omega_n^3(\omega_n+\gamma_{ij}+\gamma_{ji})^4}
\,,\nonumber \\
c_{ij}&=\pi T N_i\sum_{n=0}^{N_d}
\frac{  \gamma_{ij}(\omega_n+\gamma_{ji})(\omega_n+\gamma_{ij})
(\gamma_{ij}+\gamma_{ji}) }
{\omega_n^3(\omega_n+\gamma_{ij}+\gamma_{ji})^4}
\,.
}

In Eq.~\Eqref{Eq:GLparameters:K}, the coefficients of the gradient terms 
depend on  electronic diffusivity coefficients $D_1$ and $D_2$. The parameter 
space can be reduced by absorbing one of the electronic diffusivity coefficients 
in the gradient term. Without loss of generality, we choose $D_1$ to be the 
largest diffusivity coefficient ($D_1>D_2$). Thus, in the dimensionless units, 
the coefficients of the gradient term depend only on the ratio of diffusivities, 
or \emph{diffusivity imbalance} $r_d=D_2/D_1<1$. We define the dimensionless 
variables by normalizing the gaps by $T_c$ and the lengths by 
$\xi_0 = \sqrt{D_1/T_{c}}$. The magnetic field by is scaled 
$B_0 = T_c \sqrt{4\pi N_1}$, where $N_1$ is the density of states in the first 
band, and the free energy ${\F_0} =B_0^2/4\pi $. The electromagnetic coupling 
constant is $q = 2\pi B_0 \xi_0^2/\Phi_0$. In these units, the London penetration 
length $\lambda_L$ is given by 
$\lambda^{-2}_L=q^2(k_{ii}\Delta_{i0}^2+2 k_{12}\Delta_{10}\Delta_{20})$,
where $\Delta_{i0}$ is the bulk value of the dimensionless gap. 

\begin{figure}[!tb]
\hbox to \linewidth{ \hss
\includegraphics[width=0.99\linewidth]{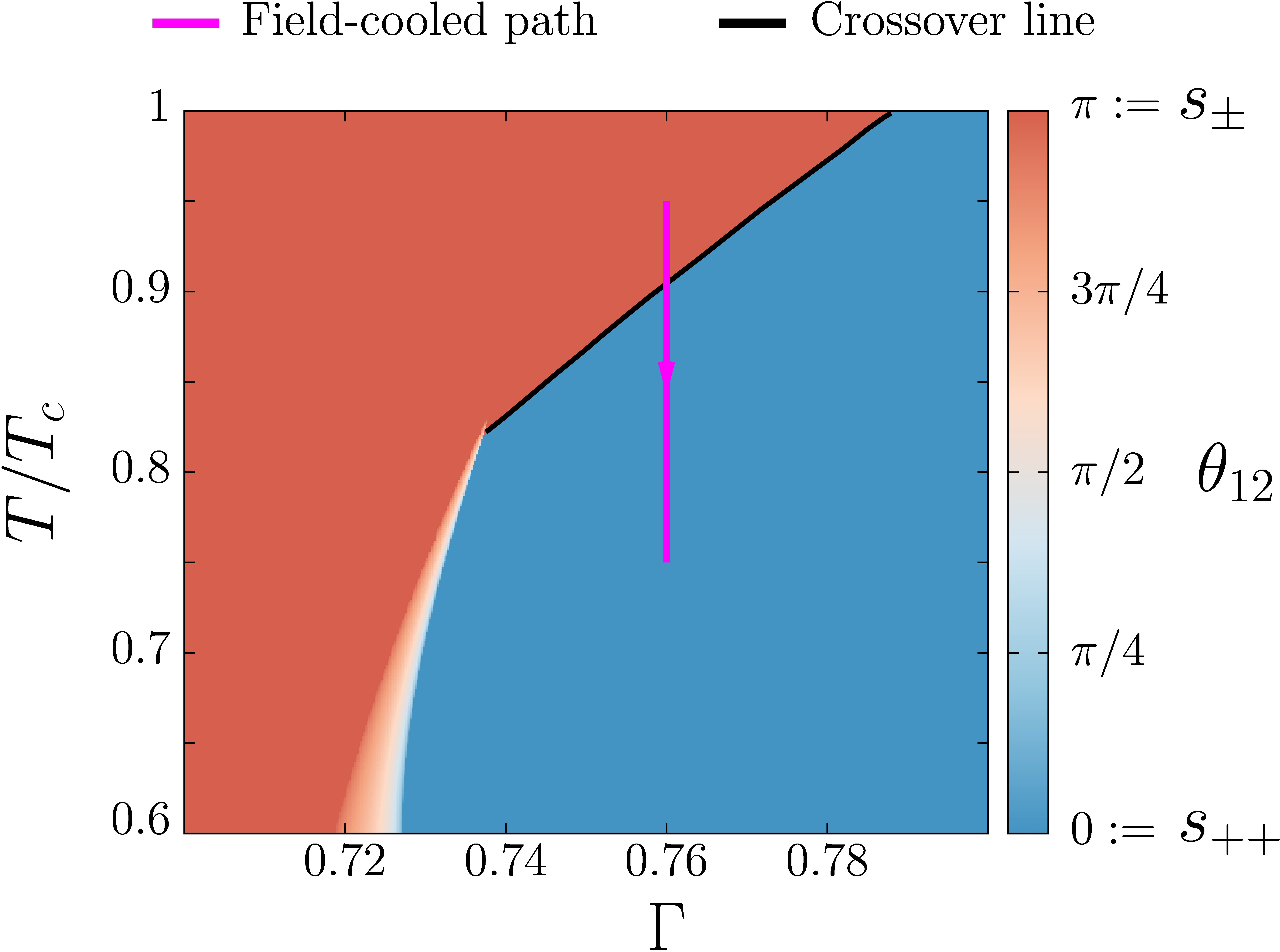}
\hss}
\caption{
Example of a phase diagram of the Ginzburg-Landau free energy describing 
two-band superconductors with interband impurity scattering. It shows the 
values of the lowest-energy state relative phase $\theta_{12}=\theta_2-\theta_1$ 
between the two components of the order parameter, as a function of temperature 
and interband scattering amplitude $\Gamma$. 
The coupling matrix $\hat{\lambda}$ corresponds to $\lambda_{11}=0.29$ and 
$\lambda_{22}=0.3$ with intermediate $\lambda_{12}=\lambda_{21}=-0.05$ repulsive 
interband pairing interaction. The solid black line shows the zero of the 
subdominant gap $\Delta_1$, which is the crossover between $s_\pm$ and $s_{++}$ 
states. The vertical line shows a field-cooling path realized later for a 
simulation in the external field.
}
\label{Fig:Diagram}
\end{figure}

It was demonstrated in Ref.~\onlinecite{Silaev.Garaud.ea:17} that, within 
its range of applicability, the two-band Ginzburg-Landau formalism 
\Eqref{Eq:FreeEnergy} indeed produces phase diagrams that match those of the 
microscopic theory even at temperatures substantially lower than the critical 
temperature $T_c$ of the superconducting phase transition.
Figure \ref{Fig:Diagram} shows such a phase diagram in the case of a two-band 
superconductor with nearly degenerate bands ($\lambda_{11}=0.29$ and 
$\lambda_{22}=0.3$), and intermediate repulsive interband pairing interaction 
($\lambda_{12}=\lambda_{21}=-0.05$). The regions of different ground state 
relative phases clearly identify the different phases. This illustrates that the 
presence of disorder leads to a crossover from the $s_{\pm}$ to the $s_{++}$ 
state. In the absence of an external magnetic field, the crossover is sharp as 
can be seen by the solid black line that shows the crossover between $s_\pm$ 
and $s_{++}$ states, where the subdominant gap $\Delta_1$ vanishes.

Another interesting transition between the $s_{++}$ and the $s_\pm$ phases 
is possible. Indeed, as a results of the presence of impurities, a possibility 
for an additional phase appears. This extra phase, where the interband phase 
difference $\theta_{12}=\theta_2-\theta_1$ is neither zero nor $\pi$, is termed 
the $s+is$ state. The fact that the interband phase difference can be such that 
$\theta_{12}\neq0,\pi$ follows from the existence of phase-locking terms 
$\propto\cos\theta$ in \Eqref{Eq:FreeEnergy:Interaction1} and the others that 
are $\propto\cos2\theta$ in \Eqref{Eq:FreeEnergy:Interaction1}. The competition 
between those terms is responsible for the existence of the impurity-induced 
$s+is$ state \cite{Bobkov.Bobkova:11}. The $s$-wave states that spontaneously 
break the time-reversal symmetry have been a subject of much interest in  
recent years; see, e.g., \cite{Ng.Nagaosa:09,Stanev.Tesanovic:10,Carlstrom.Garaud.ea:11a,
Maiti.Chubukov:13,Bobkov.Bobkova:11,Stanev.Koshelev:14,Silaev.Garaud.ea:17,
Boeker.Volkov.ea:17,Grinenko.Materne.ea:17}.
In this work, we focus on the physics of the direct $s_{\pm}$ to the $s_{++}$ 
impurity-induced crossover in an external magnetic field (denoted by the solid 
black line on \Figref{Fig:Diagram}).
Although the physics of vortices in, and in the vicinity of, the $s+is$ phase is 
very rich, it is beyond the scope of the current work and its detailed study will 
be addressed in a subsequent work \cite{tobedone}.

Below, we discuss that in an external applied magnetic field, dirty two-band 
superconductors do not feature a sharp $s_{\pm}\to s_{++}$ crossover but there 
appears a finite region in the parameter space where both 
$s_{\pm}$ and $s_{++}$ states can coexist in a peculiar way.

\section{Structural transition of vortex cores}\label{Sec:Vortices}

When going from the $s_{++}$ to $s_\pm$ state, there is a transition in the 
vortex core structure: the vortices there can develop an additional circular 
nodal line around the singular point in one of the superconducting components 
\cite{Garaud.Silaev.ea:17a}. Thus these vortices consist of an $s_\pm$ phase 
inclusion in the vortex core, which is separated from the bulk $s_{++}$ phase 
by the nodal line. 

In order to investigate the properties of single-vortex solutions, the vector 
potential $\A$ and the gap functions $\Delta_{1,2}$ are discretized using a 
finite-element framework \cite{Hecht:12}. Starting with an initial configuration 
where both components have the same phase winding (i.e., at large distances 
$\Delta_i\propto\Exp{i\theta}$ where $\theta$ is the polar angle relative to 
the vortex center), the Ginzburg-Landau free energy \Eqref{Eq:FreeEnergy} 
is then minimized using a non-linear conjugate gradient algorithm. 
We begin by simulating the vortex solutions in zero external field. 
For that purpose, the vortices are induced only by the initial configuration 
of the phase winding. For further details on the numerical methods employed 
here, see for example the related discussion in \cite{Garaud.Babaev.ea:16}.

\begin{figure}[!htb]
\hbox to \linewidth{ \hss
\includegraphics[width=\linewidth]{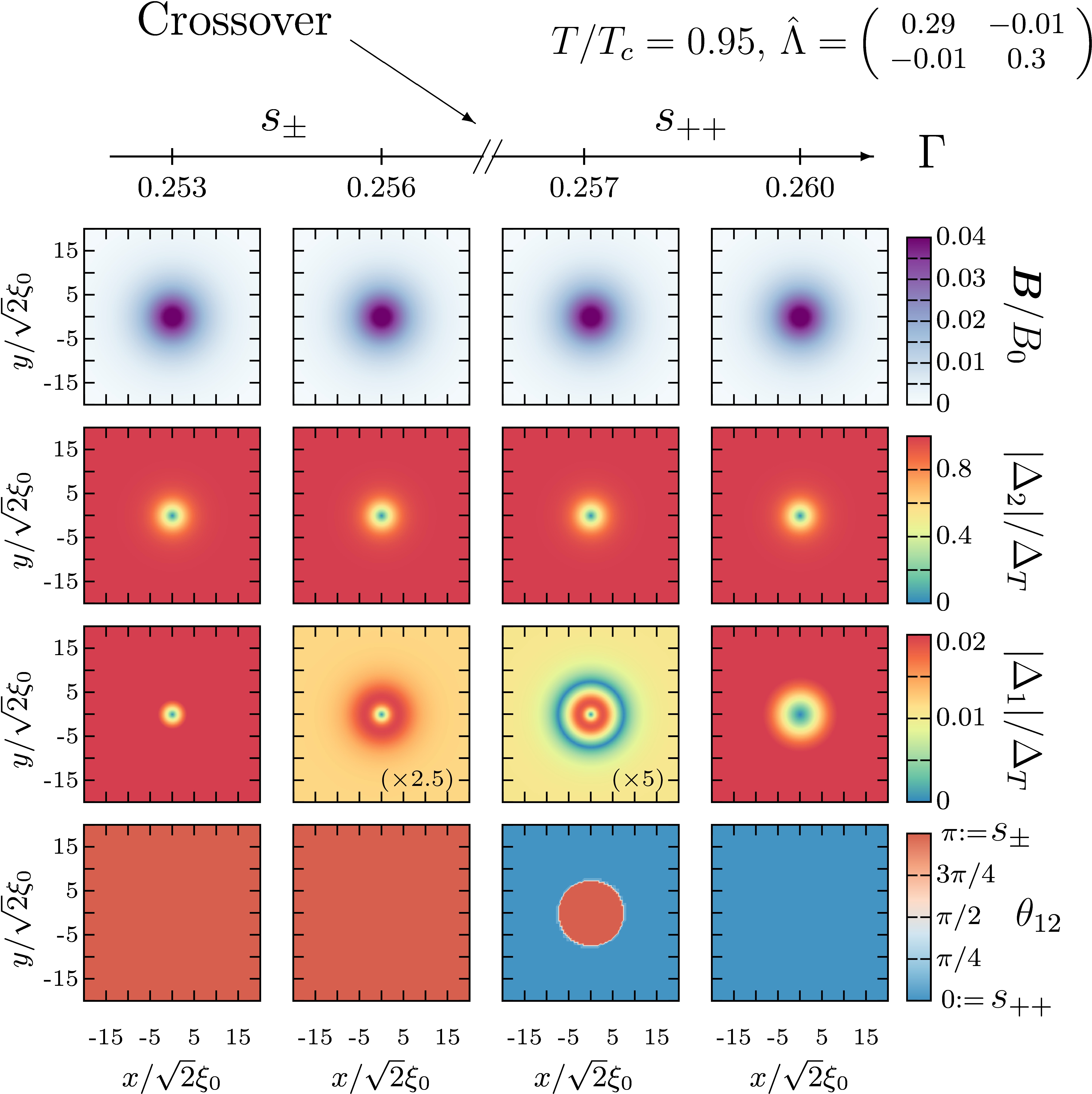}
\hss}
\caption{
Vortex configurations in the vicinity of the impurity induced crossover line 
of a two-band superconductor. The temperature here is $T/T_c=0.95$, $q=0.5$, 
and tuning the strength of the interband impurity scattering drives the ground 
state from bulk $s_{++}$ to bulk $s_\pm$. 
The different lines respectively display the magnetic field $\B$ and 
the majority ($\Delta_2$) and minority ($\Delta_1$) gap components. 
The last line shows the relative phase $\theta_{12}$ that specifies 
whether the superconducting ground state is $s_{++}$ or $s_\pm$.
In the vicinity of the impurity induced crossover line, the minority 
component of the order parameter is small, accounting for a few percent 
of the total density. 
The third column shows a vortex solution that features, in addition to the 
usual point singularities, a circular nodal line in the minority component 
$\Delta_1$.
}
\label{Fig:Vortices}
\end{figure}

Figure~\ref{Fig:Vortices} shows the numerically calculated (isolated) vortex 
solutions in the vicinity of the impurity-induced crossover, in the case of 
a two-band superconductor with nearly degenerate bands and weak repulsive 
interband pairing interaction.
Deep in the $s_{++}$ and $s_\pm$ regimes (in the first and fourth column), 
the vortices have multicore structure where both components exhibit a vortex 
profile with different sizes of conventional cores, determined by the fundamental 
length scales.
The vortex profiles of the minority component $\Delta_1$ become very different 
when approaching the crossover line. On the $s_\pm$ side of the crossover, the 
minority component exhibits a strong overshoot near the core. The density overshoot 
effect, although much smaller, was also reported in the microscopic model with 
one clean and one dirty band \cite{Tanaka.Eschrig.ea:07}.

\begin{figure}[!tb]
\hbox to \linewidth{ \hss
\includegraphics[width=\linewidth]{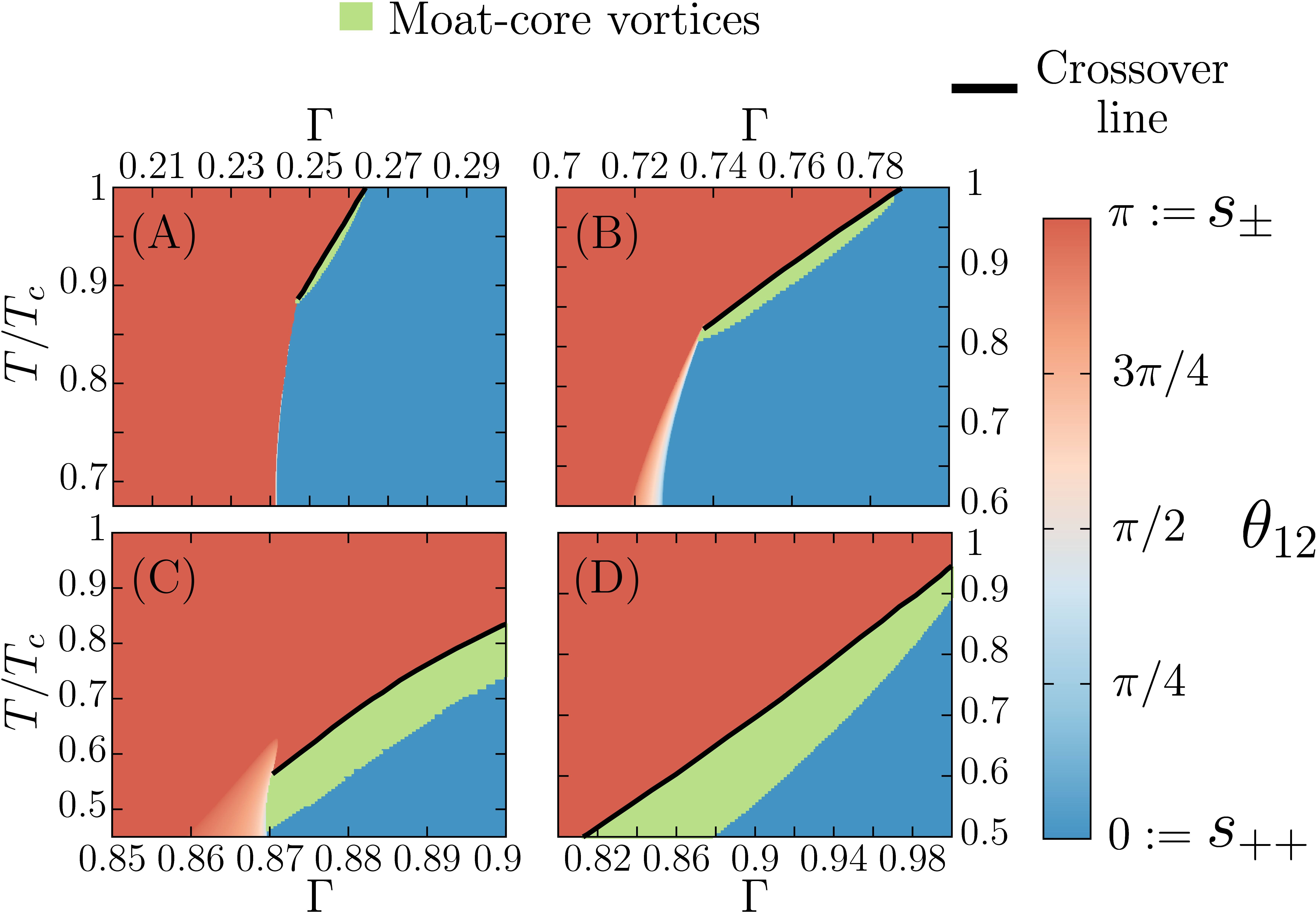}
\hss}
\caption{
Phase diagrams of the Ginzburg-Landau free energy \Eqref{Eq:FreeEnergy} 
describing two-band superconductors with interband impurity scattering. 
These show the values of the lowest-energy state relative phase 
$\theta_{12}=\theta_2-\theta_1$ between the components of the order 
parameter, and the regions of existence of moat-core vortices, as functions 
of the temperature and interband scattering $\Gamma$. 
In addition to the ground state properties, these diagrams shows the domains 
of existence of (isolated) moat-core vortices that feature a nodal line 
singularity surrounding the point singularity of the core.
The different panels correspond to different values of the coupling 
matrix $\hat{\lambda}$. Panels (a), (b), and (c) respectivelly correspond 
to nearly degenerate bands with $\lambda_{11}=0.29$ and $\lambda_{22}=0.3$ 
with weak $\lambda_{12}=\lambda_{21}=-0.01$, intermediate 
$\lambda_{12}=\lambda_{21}=-0.05$, and strong $\lambda_{12}=\lambda_{21}=-0.1$
repulsive interband pairing interaction. The last panel (d) describes the 
case of intermediate band disparity $\lambda_{11}=0.25$ and $\lambda_{22}=0.3$ 
with intermediate $\lambda_{12}=\lambda_{21}=-0.05$ repulsive interband 
pairing interaction. The solid black line shows the zero of $\Delta_2$, 
which is the crossover between $s_\pm$ and $s_{++}$ states. 
It is clear here that (isolated) vortices with nodal zero line are 
quite generic solutions in the vicinity of the crossover line.
}
\label{Fig:MoatDiagram}
\end{figure}

The very unconventional feature of vortices in this model of dirty two-band 
superconductors is the appearance of an additional circular nodal line of the 
minority component in addition to the usual point singularity at the center of 
the vortex \cite{Garaud.Silaev.ea:17a}. While the bulk relative phase is zero 
(the $s_{++}$ state) far from the vortex center, due to the competition between 
gradient and potential terms, it is more favorable to achieve a $\theta_{12}=\pi$ 
relative phase ($s_{\pm}$ state) in the vicinity of the core singularity. 
The transition between the ``core" states with $\theta_{12}=\pi$ and the 
asymptotic state $\theta_{12}=0$ is realized by nullifying the minority component 
at a given distance of the core, when the potential terms dominate over the 
gradient term. This can be seen in the third column of \Figref{Fig:Vortices}. 
Analytical explanation of such behavior was given in the previous work 
Ref.~\cite{Garaud.Silaev.ea:17a}. 
The existence of the circular nodal line results in the formation of a peculiar 
``moat"-like profile in the subdominant component of the superconducting gap.
Note that the total superconducting order parameter does not vanish there, 
rather it is only the subdominant gap (here $\Delta_1$) that vanishes exactly at 
this nodal line. Indeed, since it separates regions with $\theta_{12}=\pi$ from 
regions with zero phase difference, it has to be exactly zero somewhere. 
This can be viewed as a real-space counterpart of the crossover on the phase diagram. 
Namely, the line of direct crossover that separates the $s_{++}$ state from the 
$s_\pm$ phase is that where the subdominant gap vanishes.

\begin{figure*}[!htb]
\hbox to \linewidth{ \hss
\includegraphics[width=0.95\linewidth]{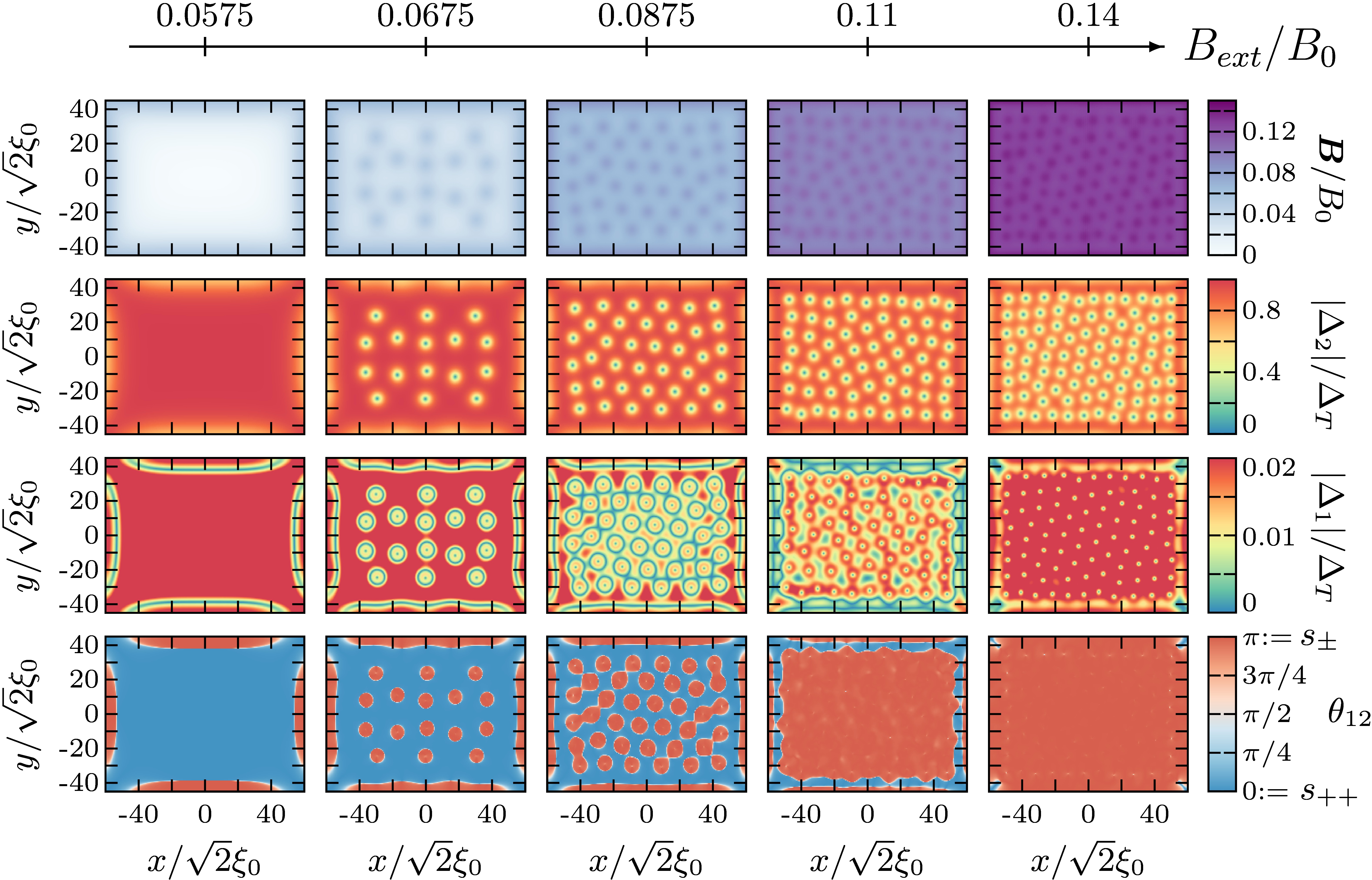}
\hss}
\caption{
Magnetization process of a dirty two-band superconductor. The coupling 
parameters are those of \Figref{Fig:Vortices}, the temperature is $T/T_c=0.9$, 
$q=0.5$, and the strength of the interband impurity scattering $\Gamma=0.7625$ 
places the system in the bulk $s_{++}$ state, in the vicinity of the crossover 
line. That choice of parameters gives single-vortex solutions with the 
circular nodal line similar to that of the third column in 
\Figref{Fig:Vortices}.
The different lines respectively display the magnetic field $\B$ and 
the majority ($\Delta_2$) and minority ($\Delta_1$) gap components. 
The last line shows the relative phase $\theta_{12}$ that specifies 
whether the superconducting ground state is $s_{++}$ or $s_\pm$.
The preferred phase locking in the bulk is $\theta_{12}=0$ (the $s_{++}$ 
state). Upon increasing external field, vortices start to enter the 
system, introducing small blobs of $\theta_{12}=\pi$ in their moat core. 
When the density of vortices becomes significant, the cores of the 
subdominant component start to overlap. As can be seen in the third column,
at intermediate vortex densities there appear regions where the blobs
of $\theta_{12}=\pi$ phase merge. When the field is increased further
the whole system shows a $\theta_{12}=\pi$ phase-locking everywhere.
Note that on the last line the system has the $s_{++}$ state in the 
bulk but $s_\pm$ state near the boundary. By contrast at elevated fields
the system becomes an $s_\pm$ superconductor in the bulk, while a 
layer of $s_{++}$ forms close to a boundary.
}
\label{Fig:Magnetization}
\end{figure*}

In order to understand how generic are such solutions, we further investigate 
numerically their existence on various phase diagrams. We find that the existence 
of these kinds of ``moat-core" vortices does not depend on the specific values of 
the pairing coefficients. 
By that statement we mean that, for the various different coupling matrices 
$\hat{\lambda}$ we investigated, we have been able to identify regions of the 
$(T,\Gamma)$ diagram where ``moat-core" vortices do form. This can be heuristically 
addressed with the following criterion that the condition for the transition of the
vortex core structure is that the mixed-gradient energy \Eqref{Eq:FreeEnergy:Mixed} 
dominates the Josephson energy \Eqref{Eq:FreeEnergy:Interaction1}. This was 
discussed analytically in Ref.~\onlinecite{Garaud.Silaev.ea:17a}.
The investigation of the vortex solutions, for various parameter sets, rather 
shows that the moat-core is a common feature in the vicinity of the crossover 
line. The region of their existence is shown in \Figref{Fig:MoatDiagram}.
It is is clearly visible that the region with moat-core vortex solutions shrinks 
close to $T_c$ and is eventually suppressed [panel (a) and (b)]. This is to be 
expected because near critical temperature, the solutions are dominated by a 
critical mode (see corresponding analytical estimates in ~\cite{Garaud.Silaev.ea:17a}). 
For the investigated regimes, the width (in terms of impurities $\delta\Gamma$) 
of the regions where moat-core vortices exist increases with increasing the 
band disparity $\lambda_{ii}$ [compare for example panels (b) and (d)]. 
By comparing panels (a), (b) and (c), it can also be seen that the width of
the moat ``region'' increases with the interband coupling $\lambda_{12}$.

\section{Phase coexistence in external field and magnetic-field-driven crossover}
\label{Sec:Magnetization}

In this section, we consider the effect in the presence of intervortex
interactions and its implication for states of dirty two-band superconductors 
in an external magnetic field.
First of all the existence of these moat-core vortices, where both $s_\pm$ and 
$s_{++}$ phases coexist, implies that a lattice of such vortices would represent 
a special kind of phase coexistence and that, in an external field, the sharp 
crossover found in the ground state is rather washed out to a finite region 
in the parameter space. 

\begin{figure*}[!htb]
\hbox to \linewidth{ \hss
\includegraphics[width=0.95\linewidth]{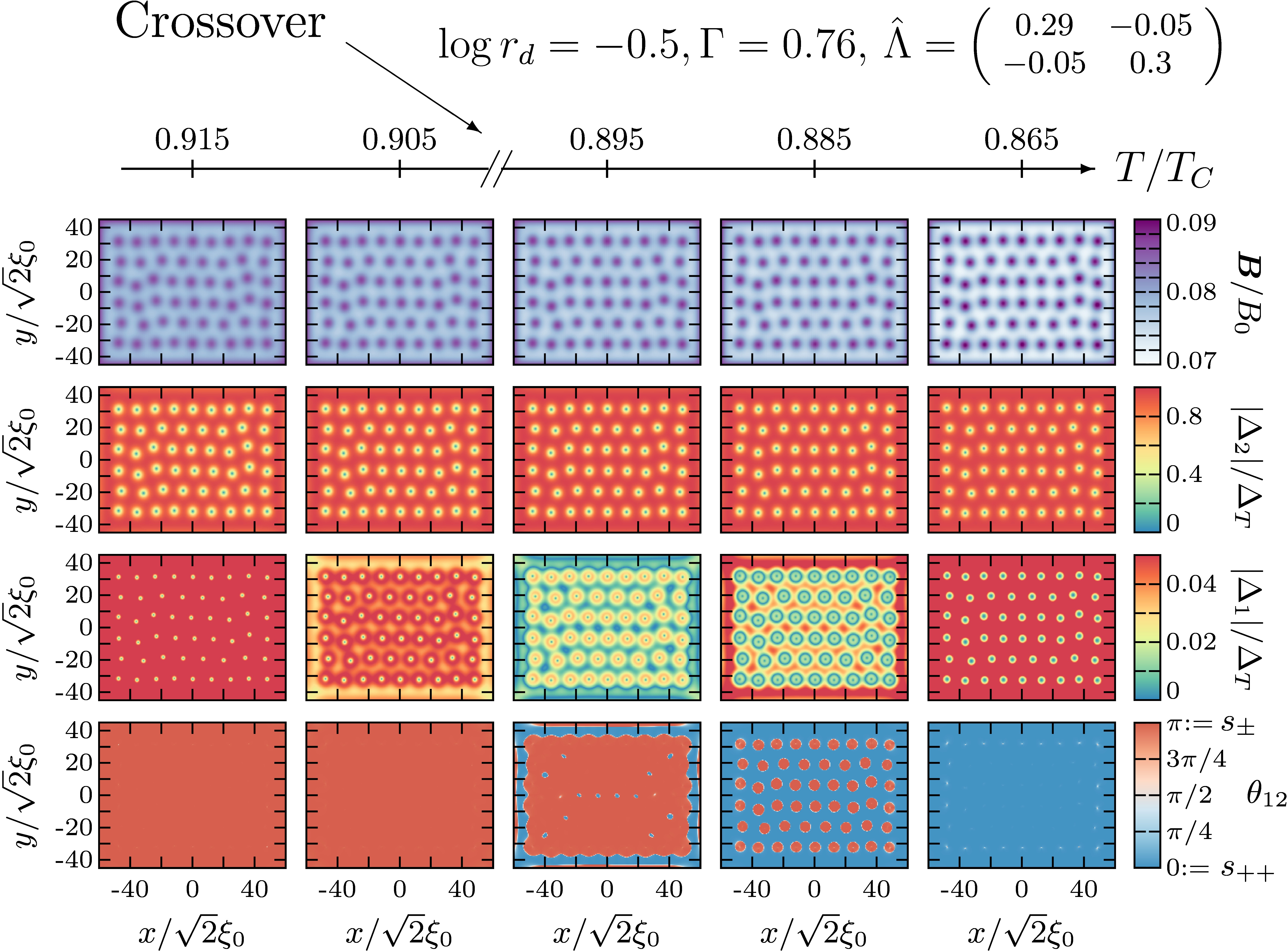}
\hss}
\caption{
A field-cooled state simulation for a dirty two-band superconductor, 
in the vicinity of the impurity induced crossover line. The displayed 
quantities are the same as in \Figref{Fig:Magnetization}.
The preferred phase locking at high temperatures is $\theta_{12}=\pi$ 
(the $s_\pm$ state), while it is $\theta_{12}=0$ (the $s_{++}$ state)
at low temperatures. Decreasing the temperatures drives the system 
across the crossover line. After the crossover, vortices still carry 
inclusions of $\theta_{12}=\pi$ in their core. Only far below  
the crossover temperature, vortices do not carry different phase locking 
in their core anymore, and the whole system shows a $\theta_{12}=0$ 
phase locking everywhere.
}
\label{Fig:Field-Cooled}
\end{figure*}

In order to investigate the response to an applied external magnetic 
field ${\bs H}=H_z{\bs e}_z$, perpendicular to the plane, the Gibbs 
free energy ${\mathcal G}=\F-\B\cdot{\bs H}$ is minimized, with 
requiring that $\Curl\A={\bs B}_{ext}$ on the boundary (for details, 
see a related discussion in Ref.~\onlinecite{Garaud.Babaev.ea:16}).

Figure~\ref{Fig:Magnetization} demonstrates the external-magnetic-field-driven 
crossover between the $s_{\pm}$ and $s_{++}$ states. The coupling parameters 
are those of \Figref{Fig:Vortices}, and the temperature and the strength of 
the interband impurity scattering place the system in the bulk $s_{++}$ state, 
close to the crossover line. The computations are thus performed for parameters 
where the single-vortex solutions have a circular nodal line similar to that 
shown in the third column in \Figref{Fig:Vortices}.
In zero and low external fields, the preferred phase locking in the bulk is 
$\theta_{12}=0$ (the $s_{++}$ state). Upon increasing the external field, 
vortices start to enter the domain, introducing small inclusions of state 
with $\theta_{12}=\pi$ in their core. When the density of vortices becomes 
significant, the cores of the subdominant component start to overlap until the 
whole system shows the $\theta_{12}=\pi$ phase locking everywhere. Thus the 
crossover here is driven by the external magnetic field.

We find that even in a low applied field, below the first critical field, 
the Meissner state also exhibits the unusual property of coexistence of 
both $s_{++}$ and $s_\pm$ states. Indeed, the $s_\pm$ state can be realized 
near the  boundaries while in the  bulk, it is the $s_{++}$ state. That is, 
as can be seen from the first column of \Figref{Fig:Magnetization}, the 
region carrying Meissner currents shows a $\theta_{12}=\pi$ phase locking 
(thus an $s_\pm$ state near boundaries), while the bulk is in the $s_{++}$ 
state (with $\theta_{12}=0$). This current-carrying region with $\theta_{12}=\pi$ 
is separated from the $\theta_{12}=0$ bulk by a nodal line of the subdominant 
component. That nodal line originates in the competition between phase locking 
kinetic terms favoring a $\theta_{12}=\pi$ relative phase in the current-carrying 
regions, and potential terms that favor zero phase difference. The reversal of 
the phase-locking in the current-carrying region is thus in some sense similar 
to that occurring in vortex cores in the vicinity of the crossover 
\cite{Garaud.Silaev.ea:17a}.

As illustrated in \Figref{Fig:Magnetization}, the magnetization properties 
in the vicinity of the $s_\pm /s_{++}$ crossover are rather unusual. 
It is also interesting to consider the case of a field-cooled simulation 
close to the impurity-induced crossover. Figure~\ref{Fig:Field-Cooled} 
displays a simulation of a field-cooled experiment at a constant applied 
field that starts at temperatures placing the system above the crossover 
line in the $s_\pm$ phase, and that ends in the $s_{++}$ phase. This 
field-cooled path is denoted by the vertical line displayed in the phase 
diagram of \Figref{Fig:Diagram}.
Sufficiently far from the crossover, the relative phase corresponds to that 
of the bulk properties in zero field, i.e., the $s_\pm$ phase for temperatures 
above the crossover line and $s_{++}$ below the crossover. However, unlike 
the zero field case, there is no sharp crossover between both states. 
As can be seen in the third column, after the crossover, the vortex-carrying 
regions remain mostly with the $s_\pm$ phase because the cores of the subdominant 
component still overlap. When going away from the crossover, that is, when 
further decreasing the temperature, the cores in the subdominant component 
$\Delta_1$ do not overlap anymore. Still the vortices with zero nodal line 
feature inclusion of the $\theta_{12}=\pi$ state while the intervortex spaces have 
relative phase $\theta_{12}=0$. That situation of a lattice (or liquid) of 
moat-core vortices thus represents a phase  coexistence or a 
mircoemulsion of $s_\pm$ inclusions inside the $s_{++}$ state.

\section{Conclusion}
\label{Sec:Conclusion}

We considered a two-band superconducting system that has an impurity-driven 
$s_\pm/s_{++}$ crossover line. We generalize the zero-field phase diagram to 
the case where the system is subjected to external magnetic field. We report 
that it is a rather generic feature that on the $s_{++}$ side of this crossover, 
the vortex solutions are unconventional, featuring  a circular nodal line that 
segregates the $s_\pm$ phase inclusions in the cores from the bulk $s_{++}$ phase. 
Further, we demonstrated that as a consequence of such vortex solutions, 
the behavior of dirty two-band superconductors in the vicinity of the crossover 
is also rich in an external field. Indeed, in contrast to the zero-field picture 
of a sharp crossover, the lattice and liquids of moat-core vortices represent 
a lattice or a ``mircoemulsion" of $s_\pm$ inclusions inside the $s_{++}$ state. 
Moreover, as the vortex density raises in increasing field, there is also a 
field-induced crossover from $s_{++}$ to the $s_\pm$, which can be resolved 
in local phase-sensitive probes \cite{hirschfeld2015robust,Du.Yang.ea:17,
Altenfeld.Hirschfeld.ea:17}. 
We also pointed out that in these systems in an applied external field, the 
superconducting state near the Meissner current carrying boundary can be $s_\pm$ 
while it is $s_{++}$ in the bulk. This result has direct implications for 
local probes of superconducting states, such as the one proposed in
Refs.~\cite{hirschfeld2015robust,Du.Yang.ea:17,Altenfeld.Hirschfeld.ea:17}, and 
for Josephson junction experiments. The coexistence state will manifest itself 
in the existence of signatures of both the $s_\pm$ and  $s_{++}$ states depending 
on the probe's position.
The formation of nodal lines was also found in a different phenomenological 
clean three-band model where it arises for a different reason due to frustrated 
Josephson terms \cite{Carlstrom.Garaud.ea:11a} rather than originating from 
competition between mixed-gradient and potential terms. That suggests that the 
phenomenon is rather generic and applies also to other models 
featuring a crossover from the $s_\pm$ to $s_{++}$ state.

\begin{acknowledgments}
The work was supported by Swedish Research Council Grants
No.~642-2013-7837, and No.~VR2016-06122 and the Goran Gustafsson Foundation 
for Research in Natural Sciences and Medicine. M.S. was supported by 
the Academy of Finland (Project No. 297439).
The computations were performed on resources provided by the Swedish 
National Infrastructure for Computing (SNIC) at National Supercomputer 
Center at Link\"oping, Sweden.
\end{acknowledgments}

\appendix
\renewcommand{\theequation}{\Alph{section}\arabic{equation}}

\section{Ginzburg-Landau coefficients}
\label{App:Coefficients}

Table \ref{Tab:GL-coefficents} shows the actual values of the coefficients 
entering the Ginzburg-Landau functional that were used for the various 
simulations throughout the paper. 

\newcolumntype{P}[1]{>{\centering\arraybackslash}p{#1}}
\newcommand{\COLSIZ}{0.0675}
\begin{table*}[!htb]
  \centering
 \begin{tabular} 
{|P{0.05\linewidth}||P{0.115\linewidth}||
P{0.075\linewidth}|P{\COLSIZ\linewidth}||
P{\COLSIZ\linewidth}|P{\COLSIZ\linewidth}|P{\COLSIZ\linewidth}||
P{0.075\linewidth}|P{\COLSIZ\linewidth}||
P{0.075\linewidth}|P{\COLSIZ\linewidth}||
P{\COLSIZ\linewidth}|
}\hline 
	 	& Control 			& $k_{11}=k_{22}$ 	& $k_{12}$ 
							& $a_{11}$ 			& $a_{22}$ 	& $a_{12}$
							& $b_{11}=b_{22}$ 	& $b_{12}$							
							& $c_{11}=c_{22}$ 	& $c_{12}$							
							& $J_{eff}$  				\\
 		& parameter			& $(\times10^{-1})$	& $(\times10^{-1})$
 							& $(\times10^{-2})$	& $(\times10^{-2})$	& $(\times10^{-3})$
 							& $(\times10^{-1})$	& $(\times10^{-3})$
 							&$(\times10^{-2})$	& $(\times10^{-3})$
 							& $(\times10^{-2})$				\\
 \hline \hline %
 							& $\Gamma=0.253$	
 							& 4.57854			& 0.373521
 							& 3.46285			& -2.29089			& -0.1615
 							& 0.741321			& -2.80412
 							& 0.329015			& 0.3137
 							& 0.17184						\\
 
 							& $\Gamma=0.256$	
 							& 4.58081			& 0.37838
 							& 3.46165			& -2.29209			& -0.86902
 							& 0.742242			& -2.83503
 							& 0.333393			& 0.321875
 							& 0.03220						\\
Fig.~\ref{Fig:Vortices}	& $\Gamma=0.257$	
							& 4.58157			& 0.380002
 							& 3.46123			& -2.29251			& -1.10501
 							& 0.742549			& -2.84532
 							& 0.334855			& 0.324627
 							& -0.01427						\\
 
 							& $\Gamma=0.260$	
 							& 4.58386			& 0.384875
 							& 3.45997			& -2.29377			& -1.81332
 							& 0.743474			& -2.87614
 							& 0.339251			& 0.332965
 							& -0.15349						\\
\hline  \hline 
Fig.~\ref{Fig:Magnetization}&	$\B/B_0$		
							&	4.45512			& 1.06235
 							&	8.70857			& 2.95483			& -106.676
 							&	0.721483		& -4.72472
 							&	0.884065		& 2.48617
 							& 	-19.6283					\\
\hline \hline  
 							& $T/T_c=0.915$					
 							&	8.94251			& 3.90581
 							&	3.40222			& -2.51494			&-4.11727
 							&	2.89522			& -3.8533
 							&	6.31571			& 32.5610
 							& 	0.29833						\\
 
							& $T/T_c=0.905$					
							&	9.02277			& 3.96752
 							&	3.06348			& -2.85368 			&-6.22441
 							&	2.94601			& -3.44763
 							&	6.47026			& 33.5801
 							& 	0.00939						\\
Fig.~\ref{Fig:Field-Cooled}&$T/T_c=0.895$					
							&	9.10466			& 4.03077
 							&	2.72175			& -3.19541			&-8.36281
 							&	2.99822			& -3.01249
 							&	6.63016			& 34.6403
 							& 	-0.28168					\\
 
 							& $T/T_c=0.885$					
 							&	9.18824			& 4.09562
 							&	2.37699			& -3.54017			&-10.5332
 							&	3.05192			& -2.54593
 							&	6.79565			& 35.7439
 							& 	-0.57084					\\
 							& $T/T_c=0.865$					
 							&	9.36067			& 4.23033
 							&	1.67808			& -4.23908			&-14.9731
 							&	3.16401			& -1.51015
 							&	7.14448			& 38.0905
 							& 	-1.12672					\\
\hline  \hline   
  \end{tabular}
  \caption{
  Coefficients of the Ginzburg-Landau free energy functional \Eqref{Eq:FreeEnergy},  
  that correspond to the various numerical simulations reported in the main body 
  of the text. They are calculated using the formulas \Eqref{Eq:GLparameters:K}, 
  \Eqref{Eq:GLparameters:A}, \Eqref{Eq:GLparameters:B}, and 
  \Eqref{Eq:GLparameters:C}, derived consistently from the microscopic Usadel 
  theory \Eqref{Eq:Usadel}. The last column shows the effective Josephson coupling
  $J_{eff}=2(a_{12}+c_{11}|\Delta_1|^2+c_{22}|\Delta_2|^2)$. Positive (resp. 
  negative) values of $J_{eff}$ denote the $s_\pm$ (resp. $s_{++}$) phase, 
  while $J_{eff}=0$ exactly at the crossover.
}
  \label{Tab:GL-coefficents}
\end{table*}


\begin{thebibliography}{24}%
\makeatletter
\providecommand \@ifxundefined [1]{%
 \@ifx{#1\undefined}
}%
\providecommand \@ifnum [1]{%
 \ifnum #1\expandafter \@firstoftwo
 \else \expandafter \@secondoftwo
 \fi
}%
\providecommand \@ifx [1]{%
 \ifx #1\expandafter \@firstoftwo
 \else \expandafter \@secondoftwo
 \fi
}%
\providecommand \natexlab [1]{#1}%
\providecommand \enquote  [1]{``#1''}%
\providecommand \bibnamefont  [1]{#1}%
\providecommand \bibfnamefont [1]{#1}%
\providecommand \citenamefont [1]{#1}%
\providecommand \href@noop [0]{\@secondoftwo}%
\providecommand \href [0]{\begingroup \@sanitize@url \@href}%
\providecommand \@href[1]{\@@startlink{#1}\@@href}%
\providecommand \@@href[1]{\endgroup#1\@@endlink}%
\providecommand \@sanitize@url [0]{\catcode `\\12\catcode `\$12\catcode
  `\&12\catcode `\#12\catcode `\^12\catcode `\_12\catcode `\%12\relax}%
\providecommand \@@startlink[1]{}%
\providecommand \@@endlink[0]{}%
\providecommand \url  [0]{\begingroup\@sanitize@url \@url }%
\providecommand \@url [1]{\endgroup\@href {#1}{\urlprefix }}%
\providecommand \urlprefix  [0]{URL }%
\providecommand \Eprint [0]{\href }%
\providecommand \doibase [0]{http://dx.doi.org/}%
\providecommand \selectlanguage [0]{\@gobble}%
\providecommand \bibinfo  [0]{\@secondoftwo}%
\providecommand \bibfield  [0]{\@secondoftwo}%
\providecommand \translation [1]{[#1]}%
\providecommand \BibitemOpen [0]{}%
\providecommand \bibitemStop [0]{}%
\providecommand \bibitemNoStop [0]{.\EOS\space}%
\providecommand \EOS [0]{\spacefactor3000\relax}%
\providecommand \BibitemShut  [1]{\csname bibitem#1\endcsname}%
\let\auto@bib@innerbib\@empty
\bibitem [{\citenamefont {Mazin}\ \emph {et~al.}(2008)\citenamefont {Mazin},
  \citenamefont {Singh}, \citenamefont {Johannes},\ and\ \citenamefont
  {Du}}]{Mazin.Singh.ea:08}%
  \BibitemOpen
  \bibfield  {author} {\bibinfo {author} {\bibfnamefont {I.~I.}\ \bibnamefont
  {Mazin}}, \bibinfo {author} {\bibfnamefont {D.~J.}\ \bibnamefont {Singh}},
  \bibinfo {author} {\bibfnamefont {M.~D.}\ \bibnamefont {Johannes}}, \ and\
  \bibinfo {author} {\bibfnamefont {M.~H.}\ \bibnamefont {Du}},\ }\bibfield
  {title} {\enquote {\bibinfo {title} {{Unconventional Superconductivity with a
  Sign Reversal in the Order Parameter of
  ${\mathrm{LaFeAsO}}_{1-x}{\mathrm{F}}_{x}$}},}\ }\href {\doibase
  10.1103/PhysRevLett.101.057003} {\bibfield  {journal} {\bibinfo  {journal}
  {Phys. Rev. Lett.}\ }\textbf {\bibinfo {volume} {101}},\ \bibinfo {pages}
  {057003} (\bibinfo {year} {2008})}\BibitemShut {NoStop}%
\bibitem [{\citenamefont {Chubukov}\ \emph {et~al.}(2008)\citenamefont
  {Chubukov}, \citenamefont {Efremov},\ and\ \citenamefont
  {Eremin}}]{Chubukov.Efremov.ea:08}%
  \BibitemOpen
  \bibfield  {author} {\bibinfo {author} {\bibfnamefont {A.~V.}\ \bibnamefont
  {Chubukov}}, \bibinfo {author} {\bibfnamefont {D.~V.}\ \bibnamefont
  {Efremov}}, \ and\ \bibinfo {author} {\bibfnamefont {I.}~\bibnamefont
  {Eremin}},\ }\bibfield  {title} {\enquote {\bibinfo {title} {{Magnetism,
  superconductivity, and pairing symmetry in iron-based superconductors}},}\
  }\href {\doibase 10.1103/PhysRevB.78.134512} {\bibfield  {journal} {\bibinfo
  {journal} {Phys. Rev. B}\ }\textbf {\bibinfo {volume} {78}},\ \bibinfo
  {pages} {134512} (\bibinfo {year} {2008})}\BibitemShut {NoStop}%
\bibitem [{\citenamefont {Hirschfeld}\ \emph {et~al.}(2011)\citenamefont
  {Hirschfeld}, \citenamefont {Korshunov},\ and\ \citenamefont
  {Mazin}}]{Hirschfeld.Korshunov.ea:11}%
  \BibitemOpen
  \bibfield  {author} {\bibinfo {author} {\bibfnamefont {P.~J.}\ \bibnamefont
  {Hirschfeld}}, \bibinfo {author} {\bibfnamefont {M.~M.}\ \bibnamefont
  {Korshunov}}, \ and\ \bibinfo {author} {\bibfnamefont {I.~I.}\ \bibnamefont
  {Mazin}},\ }\bibfield  {title} {\enquote {\bibinfo {title} {{Gap symmetry and
  structure of Fe-based superconductors}},}\ }\href {\doibase
  10.1088/0034-4885/74/12/124508} {\bibfield  {journal} {\bibinfo  {journal}
  {Reports on Progress in Physics}\ }\textbf {\bibinfo {volume} {74}},\
  \bibinfo {pages} {124508} (\bibinfo {year} {2011})}\BibitemShut {NoStop}%
\bibitem [{\citenamefont {Efremov}\ \emph {et~al.}(2011)\citenamefont
  {Efremov}, \citenamefont {Korshunov}, \citenamefont {Dolgov}, \citenamefont
  {Golubov},\ and\ \citenamefont {Hirschfeld}}]{Efremov.Korshunov.ea:11}%
  \BibitemOpen
  \bibfield  {author} {\bibinfo {author} {\bibfnamefont {D.~V.}\ \bibnamefont
  {Efremov}}, \bibinfo {author} {\bibfnamefont {M.~M.}\ \bibnamefont
  {Korshunov}}, \bibinfo {author} {\bibfnamefont {O.~V.}\ \bibnamefont
  {Dolgov}}, \bibinfo {author} {\bibfnamefont {A.~A.}\ \bibnamefont {Golubov}},
  \ and\ \bibinfo {author} {\bibfnamefont {P.~J.}\ \bibnamefont {Hirschfeld}},\
  }\bibfield  {title} {\enquote {\bibinfo {title} {{Disorder-induced transition
  between ${s}_{\ifmmode\pm\else\textpm\fi{}}$ and ${s}_{++}$ states in
  two-band superconductors}},}\ }\href {\doibase 10.1103/PhysRevB.84.180512}
  {\bibfield  {journal} {\bibinfo  {journal} {Phys. Rev. B}\ }\textbf {\bibinfo
  {volume} {84}},\ \bibinfo {pages} {180512} (\bibinfo {year}
  {2011})}\BibitemShut {NoStop}%
\bibitem [{\citenamefont {Garaud}\ \emph {et~al.}(2017)\citenamefont {Garaud},
  \citenamefont {Silaev},\ and\ \citenamefont {Babaev}}]{Garaud.Silaev.ea:17a}%
  \BibitemOpen
  \bibfield  {author} {\bibinfo {author} {\bibfnamefont {Julien}\ \bibnamefont
  {Garaud}}, \bibinfo {author} {\bibfnamefont {Mihail}\ \bibnamefont {Silaev}},
  \ and\ \bibinfo {author} {\bibfnamefont {Egor}\ \bibnamefont {Babaev}},\
  }\bibfield  {title} {\enquote {\bibinfo {title} {{Change of the vortex core
  structure in two-band superconductors at the impurity-scattering-driven
  $s_\pm/s_{++}$ crossover}},}\ }\href {\doibase 10.1103/PhysRevB.96.140503}
  {\bibfield  {journal} {\bibinfo  {journal} {Phys. Rev. B}\ }\textbf {\bibinfo
  {volume} {96}},\ \bibinfo {pages} {140503} (\bibinfo {year}
  {2017})}\BibitemShut {NoStop}%
\bibitem [{\citenamefont {Gurevich}(2003)}]{Gurevich:03}%
  \BibitemOpen
  \bibfield  {author} {\bibinfo {author} {\bibfnamefont {A.}~\bibnamefont
  {Gurevich}},\ }\bibfield  {title} {\enquote {\bibinfo {title} {{Enhancement
  of the upper critical field by nonmagnetic impurities in dirty two-gap
  superconductors}},}\ }\href {\doibase 10.1103/PhysRevB.67.184515} {\bibfield
  {journal} {\bibinfo  {journal} {Phys. Rev. B}\ }\textbf {\bibinfo {volume}
  {67}},\ \bibinfo {pages} {184515} (\bibinfo {year} {2003})}\BibitemShut
  {NoStop}%
\bibitem [{Note1()}]{Note1}%
  \BibitemOpen
  \bibinfo {note} {For the question of formal validity of the multiband
  expansion see the discussion of the clean case \cite
  {Silaev.Babaev:12}}\BibitemShut {NoStop}%
\bibitem [{\citenamefont {Silaev}\ and\ \citenamefont
  {Babaev}(2012)}]{Silaev.Babaev:12}%
  \BibitemOpen
  \bibfield  {author} {\bibinfo {author} {\bibfnamefont {Mihail}\ \bibnamefont
  {Silaev}}\ and\ \bibinfo {author} {\bibfnamefont {Egor}\ \bibnamefont
  {Babaev}},\ }\bibfield  {title} {\enquote {\bibinfo {title} {{Microscopic
  derivation of two-component Ginzburg-Landau model and conditions of its
  applicability in two-band systems}},}\ }\href {\doibase
  10.1103/PhysRevB.85.134514} {\bibfield  {journal} {\bibinfo  {journal} {Phys.
  Rev. B}\ }\textbf {\bibinfo {volume} {85}},\ \bibinfo {pages} {134514}
  (\bibinfo {year} {2012})}\BibitemShut {NoStop}%
\bibitem [{\citenamefont {Silaev}\ \emph {et~al.}(2017)\citenamefont {Silaev},
  \citenamefont {Garaud},\ and\ \citenamefont {Babaev}}]{Silaev.Garaud.ea:17}%
  \BibitemOpen
  \bibfield  {author} {\bibinfo {author} {\bibfnamefont {Mihail}\ \bibnamefont
  {Silaev}}, \bibinfo {author} {\bibfnamefont {Julien}\ \bibnamefont {Garaud}},
  \ and\ \bibinfo {author} {\bibfnamefont {Egor}\ \bibnamefont {Babaev}},\
  }\bibfield  {title} {\enquote {\bibinfo {title} {{Phase diagram of dirty
  two-band superconductors and observability of impurity-induced $s+is$
  state}},}\ }\href {\doibase 10.1103/PhysRevB.95.024517} {\bibfield  {journal}
  {\bibinfo  {journal} {Phys. Rev. B}\ }\textbf {\bibinfo {volume} {95}},\
  \bibinfo {pages} {024517} (\bibinfo {year} {2017})}\BibitemShut {NoStop}%
\bibitem [{\citenamefont {Bobkov}\ and\ \citenamefont
  {Bobkova}(2011)}]{Bobkov.Bobkova:11}%
  \BibitemOpen
  \bibfield  {author} {\bibinfo {author} {\bibfnamefont {A.~M.}\ \bibnamefont
  {Bobkov}}\ and\ \bibinfo {author} {\bibfnamefont {I.~V.}\ \bibnamefont
  {Bobkova}},\ }\bibfield  {title} {\enquote {\bibinfo {title} {{Time-reversal
  symmetry breaking state near the surface of an
  ${s}_{\ifmmode\pm\else\textpm\fi{}}$ superconductor}},}\ }\href {\doibase
  10.1103/PhysRevB.84.134527} {\bibfield  {journal} {\bibinfo  {journal} {Phys.
  Rev. B}\ }\textbf {\bibinfo {volume} {84}},\ \bibinfo {pages} {134527}
  (\bibinfo {year} {2011})}\BibitemShut {NoStop}%
\bibitem [{\citenamefont {Ng}\ and\ \citenamefont
  {Nagaosa}(2009)}]{Ng.Nagaosa:09}%
  \BibitemOpen
  \bibfield  {author} {\bibinfo {author} {\bibfnamefont {T.~K.}\ \bibnamefont
  {Ng}}\ and\ \bibinfo {author} {\bibfnamefont {N.}~\bibnamefont {Nagaosa}},\
  }\bibfield  {title} {\enquote {\bibinfo {title} {{Broken time-reversal
  symmetry in Josephson junction involving two-band superconductors}},}\ }\href
  {\doibase 10.1209/0295-5075/87/17003} {\bibfield  {journal} {\bibinfo
  {journal} {Europhysics Letters}\ }\textbf {\bibinfo {volume} {87}},\ \bibinfo
  {pages} {17003--+} (\bibinfo {year} {2009})}\BibitemShut {NoStop}%
\bibitem [{\citenamefont {Stanev}\ and\ \citenamefont {Te\ifmmode
  \check{s}\else \v{s}\fi{}anovi\ifmmode~\acute{c}\else
  \'{c}\fi{}}(2010)}]{Stanev.Tesanovic:10}%
  \BibitemOpen
  \bibfield  {author} {\bibinfo {author} {\bibfnamefont {Valentin}\
  \bibnamefont {Stanev}}\ and\ \bibinfo {author} {\bibfnamefont {Zlatko}\
  \bibnamefont {Te\ifmmode \check{s}\else
  \v{s}\fi{}anovi\ifmmode~\acute{c}\else \'{c}\fi{}}},\ }\bibfield  {title}
  {\enquote {\bibinfo {title} {{Three-band superconductivity and the order
  parameter that breaks time-reversal symmetry}},}\ }\href {\doibase
  10.1103/PhysRevB.81.134522} {\bibfield  {journal} {\bibinfo  {journal} {Phys.
  Rev. B}\ }\textbf {\bibinfo {volume} {81}},\ \bibinfo {pages} {134522}
  (\bibinfo {year} {2010})}\BibitemShut {NoStop}%
\bibitem [{\citenamefont {Carlstr\"om}\ \emph {et~al.}(2011)\citenamefont
  {Carlstr\"om}, \citenamefont {Garaud},\ and\ \citenamefont
  {Babaev}}]{Carlstrom.Garaud.ea:11a}%
  \BibitemOpen
  \bibfield  {author} {\bibinfo {author} {\bibfnamefont {Johan}\ \bibnamefont
  {Carlstr\"om}}, \bibinfo {author} {\bibfnamefont {Julien}\ \bibnamefont
  {Garaud}}, \ and\ \bibinfo {author} {\bibfnamefont {Egor}\ \bibnamefont
  {Babaev}},\ }\bibfield  {title} {\enquote {\bibinfo {title} {{Length scales,
  collective modes, and type-1.5 regimes in three-band superconductors}},}\
  }\href {\doibase 10.1103/PhysRevB.84.134518} {\bibfield  {journal} {\bibinfo
  {journal} {Phys. Rev. B}\ }\textbf {\bibinfo {volume} {84}},\ \bibinfo
  {pages} {134518} (\bibinfo {year} {2011})}\BibitemShut {NoStop}%
\bibitem [{\citenamefont {Maiti}\ and\ \citenamefont
  {Chubukov}(2013)}]{Maiti.Chubukov:13}%
  \BibitemOpen
  \bibfield  {author} {\bibinfo {author} {\bibfnamefont {Saurabh}\ \bibnamefont
  {Maiti}}\ and\ \bibinfo {author} {\bibfnamefont {Andrey~V.}\ \bibnamefont
  {Chubukov}},\ }\bibfield  {title} {\enquote {\bibinfo {title} {{$s+is$ state
  with broken time-reversal symmetry in {Fe}-based superconductors}},}\ }\href
  {\doibase 10.1103/PhysRevB.87.144511} {\bibfield  {journal} {\bibinfo
  {journal} {Phys. Rev. B}\ }\textbf {\bibinfo {volume} {87}},\ \bibinfo
  {pages} {144511} (\bibinfo {year} {2013})}\BibitemShut {NoStop}%
\bibitem [{\citenamefont {Stanev}\ and\ \citenamefont
  {Koshelev}(2014)}]{Stanev.Koshelev:14}%
  \BibitemOpen
  \bibfield  {author} {\bibinfo {author} {\bibfnamefont {Valentin}\
  \bibnamefont {Stanev}}\ and\ \bibinfo {author} {\bibfnamefont {Alexei~E.}\
  \bibnamefont {Koshelev}},\ }\bibfield  {title} {\enquote {\bibinfo {title}
  {{Complex state induced by impurities in multiband superconductors}},}\
  }\href {\doibase 10.1103/PhysRevB.89.100505} {\bibfield  {journal} {\bibinfo
  {journal} {Phys. Rev. B}\ }\textbf {\bibinfo {volume} {89}},\ \bibinfo
  {pages} {100505} (\bibinfo {year} {2014})}\BibitemShut {NoStop}%
\bibitem [{\citenamefont {B\"oker}\ \emph {et~al.}(2017)\citenamefont
  {B\"oker}, \citenamefont {Volkov}, \citenamefont {Efetov},\ and\
  \citenamefont {Eremin}}]{Boeker.Volkov.ea:17}%
  \BibitemOpen
  \bibfield  {author} {\bibinfo {author} {\bibfnamefont {Jakob}\ \bibnamefont
  {B\"oker}}, \bibinfo {author} {\bibfnamefont {Pavel~A.}\ \bibnamefont
  {Volkov}}, \bibinfo {author} {\bibfnamefont {Konstantin~B.}\ \bibnamefont
  {Efetov}}, \ and\ \bibinfo {author} {\bibfnamefont {Ilya}\ \bibnamefont
  {Eremin}},\ }\bibfield  {title} {\enquote {\bibinfo {title} {{$s+is$
  superconductivity with incipient bands: Doping dependence and STM
  signatures}},}\ }\href {\doibase 10.1103/PhysRevB.96.014517} {\bibfield
  {journal} {\bibinfo  {journal} {Phys. Rev. B}\ }\textbf {\bibinfo {volume}
  {96}},\ \bibinfo {pages} {014517} (\bibinfo {year} {2017})}\BibitemShut
  {NoStop}%
\bibitem [{\citenamefont {Grinenko}\ \emph {et~al.}(2017)\citenamefont
  {Grinenko}, \citenamefont {Materne}, \citenamefont {Sarkar}, \citenamefont
  {Luetkens}, \citenamefont {Kihou}, \citenamefont {Lee}, \citenamefont
  {Akhmadaliev}, \citenamefont {Efremov}, \citenamefont {Drechsler},\ and\
  \citenamefont {Klauss}}]{Grinenko.Materne.ea:17}%
  \BibitemOpen
  \bibfield  {author} {\bibinfo {author} {\bibfnamefont {V.}~\bibnamefont
  {Grinenko}}, \bibinfo {author} {\bibfnamefont {P.}~\bibnamefont {Materne}},
  \bibinfo {author} {\bibfnamefont {R.}~\bibnamefont {Sarkar}}, \bibinfo
  {author} {\bibfnamefont {H.}~\bibnamefont {Luetkens}}, \bibinfo {author}
  {\bibfnamefont {K.}~\bibnamefont {Kihou}}, \bibinfo {author} {\bibfnamefont
  {C.~H.}\ \bibnamefont {Lee}}, \bibinfo {author} {\bibfnamefont
  {S.}~\bibnamefont {Akhmadaliev}}, \bibinfo {author} {\bibfnamefont {D.~V.}\
  \bibnamefont {Efremov}}, \bibinfo {author} {\bibfnamefont {S.-L.}\
  \bibnamefont {Drechsler}}, \ and\ \bibinfo {author} {\bibfnamefont {H.-H.}\
  \bibnamefont {Klauss}},\ }\bibfield  {title} {\enquote {\bibinfo {title}
  {{Superconductivity with broken time-reversal symmetry in ion-irradiated
  Ba$_{0.27}$K$_{0.73}$Fe$_2$As$_2$ single crystals}},}\ }\href {\doibase
  10.1103/PhysRevB.95.214511} {\bibfield  {journal} {\bibinfo  {journal} {Phys.
  Rev. B}\ }\textbf {\bibinfo {volume} {95}},\ \bibinfo {pages} {214511}
  (\bibinfo {year} {2017})}\BibitemShut {NoStop}%
\bibitem [{\citenamefont {{Garaud}}\ \emph {et~al.}(2018)\citenamefont
  {{Garaud}}, \citenamefont {{Corticelli}}, \citenamefont {{Silaev}},\ and\
  \citenamefont {{Babaev}}}]{tobedone}%
  \BibitemOpen
  \bibfield  {author} {\bibinfo {author} {\bibfnamefont {J.}~\bibnamefont
  {{Garaud}}}, \bibinfo {author} {\bibfnamefont {A.}~\bibnamefont
  {{Corticelli}}}, \bibinfo {author} {\bibfnamefont {M.}~\bibnamefont
  {{Silaev}}}, \ and\ \bibinfo {author} {\bibfnamefont {E.}~\bibnamefont
  {{Babaev}}},\ }\href@noop {} {\enquote {\bibinfo {title} {{Properties of
  dirty two-bands superconductors with repulsive interband interaction: normal
  modes, length scales, vortices and magnetic response}},}\ } (\bibinfo {year}
  {2018}),\ \Eprint {http://arxiv.org/abs/1802.07252} {arXiv:1802.07252
  [cond-mat.supr-con]} \BibitemShut {NoStop}%
\bibitem [{\citenamefont {Hecht}(2012)}]{Hecht:12}%
  \BibitemOpen
  \bibfield  {author} {\bibinfo {author} {\bibfnamefont {F.}~\bibnamefont
  {Hecht}},\ }\bibfield  {title} {\enquote {\bibinfo {title} {{New development
  in Freefem++}},}\ }\href {\doibase 10.1515/jnum-2012-0013} {\bibfield
  {journal} {\bibinfo  {journal} {J. Numer. Math.}\ }\textbf {\bibinfo {volume}
  {20}},\ \bibinfo {pages} {251--265} (\bibinfo {year} {2012})}\BibitemShut
  {NoStop}%
\bibitem [{\citenamefont {Garaud}\ \emph {et~al.}(2016)\citenamefont {Garaud},
  \citenamefont {Babaev}, \citenamefont {Bojesen},\ and\ \citenamefont
  {Sudb\o{}}}]{Garaud.Babaev.ea:16}%
  \BibitemOpen
  \bibfield  {author} {\bibinfo {author} {\bibfnamefont {Julien}\ \bibnamefont
  {Garaud}}, \bibinfo {author} {\bibfnamefont {Egor}\ \bibnamefont {Babaev}},
  \bibinfo {author} {\bibfnamefont {Troels~Arnfred}\ \bibnamefont {Bojesen}}, \
  and\ \bibinfo {author} {\bibfnamefont {Asle}\ \bibnamefont {Sudb\o{}}},\
  }\bibfield  {title} {\enquote {\bibinfo {title} {{Lattices of double-quanta
  vortices and chirality inversion in ${p}_{x}+i{p}_{y}$ superconductors}},}\
  }\href {\doibase 10.1103/PhysRevB.94.104509} {\bibfield  {journal} {\bibinfo
  {journal} {Phys. Rev. B}\ }\textbf {\bibinfo {volume} {94}},\ \bibinfo
  {pages} {104509} (\bibinfo {year} {2016})}\BibitemShut {NoStop}%
\bibitem [{\citenamefont {Tanaka}\ \emph {et~al.}(2007)\citenamefont {Tanaka},
  \citenamefont {Eschrig},\ and\ \citenamefont
  {Agterberg}}]{Tanaka.Eschrig.ea:07}%
  \BibitemOpen
  \bibfield  {author} {\bibinfo {author} {\bibfnamefont {K.}~\bibnamefont
  {Tanaka}}, \bibinfo {author} {\bibfnamefont {M.}~\bibnamefont {Eschrig}}, \
  and\ \bibinfo {author} {\bibfnamefont {D.~F.}\ \bibnamefont {Agterberg}},\
  }\bibfield  {title} {\enquote {\bibinfo {title} {{Theory of vortices in
  hybridized ballistic/diffusive-band superconductors}},}\ }\href {\doibase
  10.1103/PhysRevB.75.214512} {\bibfield  {journal} {\bibinfo  {journal} {Phys.
  Rev. B}\ }\textbf {\bibinfo {volume} {75}},\ \bibinfo {pages} {214512}
  (\bibinfo {year} {2007})}\BibitemShut {NoStop}%
\bibitem [{\citenamefont {Hirschfeld}\ \emph {et~al.}(2015)\citenamefont
  {Hirschfeld}, \citenamefont {Altenfeld}, \citenamefont {Eremin},\ and\
  \citenamefont {Mazin}}]{hirschfeld2015robust}%
  \BibitemOpen
  \bibfield  {author} {\bibinfo {author} {\bibfnamefont {P.~J.}\ \bibnamefont
  {Hirschfeld}}, \bibinfo {author} {\bibfnamefont {D.}~\bibnamefont
  {Altenfeld}}, \bibinfo {author} {\bibfnamefont {I.}~\bibnamefont {Eremin}}, \
  and\ \bibinfo {author} {\bibfnamefont {I.~I.}\ \bibnamefont {Mazin}},\
  }\bibfield  {title} {\enquote {\bibinfo {title} {{Robust determination of the
  superconducting gap sign structure via quasiparticle interference}},}\ }\href
  {\doibase 10.1103/PhysRevB.92.184513} {\bibfield  {journal} {\bibinfo
  {journal} {Phys. Rev. B}\ }\textbf {\bibinfo {volume} {92}},\ \bibinfo
  {pages} {184513} (\bibinfo {year} {2015})}\BibitemShut {NoStop}%
\bibitem [{\citenamefont {Du}\ \emph {et~al.}(2017)\citenamefont {Du},
  \citenamefont {Yang}, \citenamefont {Altenfeld}, \citenamefont {Gu},
  \citenamefont {Yang}, \citenamefont {Eremin}, \citenamefont {Hirschfeld},
  \citenamefont {Mazin}, \citenamefont {Lin}, \citenamefont {Zhu},\ and\
  \citenamefont {Wen}}]{Du.Yang.ea:17}%
  \BibitemOpen
  \bibfield  {author} {\bibinfo {author} {\bibfnamefont {Zengyi}\ \bibnamefont
  {Du}}, \bibinfo {author} {\bibfnamefont {Xiong}\ \bibnamefont {Yang}},
  \bibinfo {author} {\bibfnamefont {Dustin}\ \bibnamefont {Altenfeld}},
  \bibinfo {author} {\bibfnamefont {Qiangqiang}\ \bibnamefont {Gu}}, \bibinfo
  {author} {\bibfnamefont {Huan}\ \bibnamefont {Yang}}, \bibinfo {author}
  {\bibfnamefont {Ilya}\ \bibnamefont {Eremin}}, \bibinfo {author}
  {\bibfnamefont {Peter~J.}\ \bibnamefont {Hirschfeld}}, \bibinfo {author}
  {\bibfnamefont {Igor~I.}\ \bibnamefont {Mazin}}, \bibinfo {author}
  {\bibfnamefont {Hai}\ \bibnamefont {Lin}}, \bibinfo {author} {\bibfnamefont
  {Xiyu}\ \bibnamefont {Zhu}}, \ and\ \bibinfo {author} {\bibfnamefont
  {Hai-Hu}\ \bibnamefont {Wen}},\ }\bibfield  {title} {\enquote {\bibinfo
  {title} {{Sign reversal of the order parameter in
  (Li$_{1-x}$Fe$_x$)OHFe$_{1-y}$Zn$_y$Se}},}\ }\href
  {http://dx.doi.org/10.1038/nphys4299} {\bibfield  {journal} {\bibinfo
  {journal} {Nature Physics}\ }\textbf {\bibinfo {volume} {14}},\ \bibinfo
  {pages} {134} (\bibinfo {year} {2017})}\BibitemShut {NoStop}%
\bibitem [{\citenamefont {{Altenfeld}}\ \emph {et~al.}(2017)\citenamefont
  {{Altenfeld}}, \citenamefont {{Hirschfeld}}, \citenamefont {{Mazin}},\ and\
  \citenamefont {{Eremin}}}]{Altenfeld.Hirschfeld.ea:17}%
  \BibitemOpen
  \bibfield  {author} {\bibinfo {author} {\bibfnamefont {D.}~\bibnamefont
  {{Altenfeld}}}, \bibinfo {author} {\bibfnamefont {P.~J.}\ \bibnamefont
  {{Hirschfeld}}}, \bibinfo {author} {\bibfnamefont {I.~I.}\ \bibnamefont
  {{Mazin}}}, \ and\ \bibinfo {author} {\bibfnamefont {I.}~\bibnamefont
  {{Eremin}}},\ }\bibfield  {title} {\enquote {\bibinfo {title} {{Detecting
  sign-changing superconducting gap in LiFeAs using quasiparticle
  interference}},}\ }\href@noop {} {\bibfield  {journal} {\bibinfo  {journal}
  {ArXiv e-prints}\ } (\bibinfo {year} {2017})},\ \Eprint
  {http://arxiv.org/abs/1712.03625} {arXiv:1712.03625 [cond-mat.supr-con]}
  \BibitemShut {NoStop}%
\end{thebibliography}
%

\end{document}